\documentclass[pop,12pt,fleqn]{w-art}
\usepackage{times}
\usepackage{w-thm}
\usepackage{amssymb,euscript}

\numberwithin{equation}{section}
\begin{document}
%%    The information for the title page will be placed between
%%    \begin{document} and \maketitle. The order of most entries
%%    is determined by the class file and can not be changed by
%%    rearranging them. The maketitle commaynd follows after the
%%    abstract.
%%
%%    Most of the following commands will be completed by the publisher.
%%
%%    The copyrightyear is defined in the .clo file as the first argument
%%    of the copyrightinfo command. If the copyrightyear differs from that
%%    value it might be adjusted by the following definition:
%%
%% \renewcommand{\copyrightyear}{2003}% uncomment to change the copyrightyear.
%%
\DOIsuffix{theDOIsuffix}
%%
%% issueinfo for header and copyright line
\Volume{51}
\Issue{1}
\Month{01}
\Year{2003}
%%
%%    First and last pagenumber of the article. If the option
%%    'autolastpage' is set (default) the second argument may be left empty.
\pagespan{1}{}
%%
%%    Dates will be filled in by the publisher. The 'reviseddate' and
%%    'dateposted' (Published online) entry may be left empty.
%%
%\keywords{List, of, comma, separated, keywords.}
% \subjclass[pacs]{04A25}

%% \pretitle{Editor's Choice}

%% We have a short and a long form for the title. The short form
%% (optional argument) goes into the running head.

\bigskip

\begin{flushright}
\hfill{AEI-2004-077}\\
\hfill{LMU-ASC 18/05}\\
\hfill{hep-th/0503129}\\
\hfill{March 4, 2005} \\
~ 
\end{flushright}

\title[Stabilization of moduli]{Stabilization of moduli by 
fluxes\footnote{To appear in Proceedings of: 
	``Strings and Cosmology'',
(Texas A\&M University, March 2004) and the
RTN workshop: ``The quantum structure of space-time and the geometric 
nature of fundamental interactions''
(Kolymbari, Crete, September 2004)
}
}
 
%% Please do not enter footnotes or \inst{}-notes into the optional
%% argument of the author command. The optional argument will go into
%% the header.  If there is only one address the marker \inst{x} may be
%% omitted.

%% Information for the first author.
\author[Klaus Behrndt]{Klaus Behrndt}
 \address[]{
Arnold-Sommerfeld-Center for Theoretical Physics \\
Department f\"ur Physik, Ludwig-Maximilians-Universit\"at 
M\"unchen \\
Theresienstra\ss{}e 37, 80333 M\"unchen, Germany \\[2mm]
E-mail: {\sf behrndt@theorie.physik.uni-muenchen.de}
}
%%
%%    Information for the second author

\bigskip

\begin{abstract}

  In order to lift the continuous moduli space of string vacua,
  non-trivial fluxes may be the essential input.  In this talk I
  summarize aspects of two approaches to compactifications in the
  presence of fluxes: $(i)$ generalized
  Scherk-Schwarz reductions and gauged supergravity and $(ii)$ the
  description of flux-deformed geometries in terms of $G$-structures
  and intrinsic torsion.

\end{abstract}
%% maketitle must follow the abstract.
\maketitle                   % Produces the title.

%% If there is not enough space inside the running head
%% for all authors including the title you may provide
%% the leftmark in one of the following three forms:

%% \renewcommand{\leftmark}
%% {First Author: A Short Title}

%% \renewcommand{\leftmark}
%% {First Author and Second Author: A Short Title}

%% \renewcommand{\leftmark}
%% {First Author et al.: A Short Title}

%% \tableofcontents  % Produces the table of contents.

\newcommand{\be}[3]{\begin{equation}  \label{#1#2#3}}
\newcommand{\ee}{\end{equation}}
\newcommand{\ba}{\begin{array}}
\newcommand{\ea}{\end{array}}
\newcommand{\bea}[3]{\begin{eqnarray}  \label{#1#2#3}}
\newcommand{\eea}{\end{eqnarray}}
\newcommand{\com}[1]{\begin{center} {\tt #1} \end{center}}
\newcommand{\ip}{\raise1pt\hbox{\large$\lrcorner$}\,}

\newcommand{\N}{${\cal N}\, $}

\def\I{\mathbf I}
\def\P{\mathbf P}
\def\R{\mathbf R}
\def\S{\mathbf S}
\def\X{\mathbf X}
\def\C{\mathbf C}
\def\Y{\mathbf Y}
\def\1{\mathbf 1}
\def\F{{\cal F}^{(1)}}
\def\FF{{\cal F}^{(7)}}
\def\FFF{{\cal F}^{(27)}}

\newcommand{\haken}{\mathbin{\hbox to 8pt{%
                 \vrule height0.4pt width7pt depth0pt \kern-.4pt
                 \vrule height4pt width0.4pt depth0pt\hss}}}

%%%%%%%%%%%%%%%%%%%%%%%%%%%%%%%%%%%%%%%%

\baselineskip=15pt
\parskip=3pt

%%%%%%%%%%%%%%%%%%%%%%%%%%%%%%%%%%%%%%%%%%%%%%

%%ALIAS=fluxold=hep-th/9510227,hep-th/9610151,hep-th/9912152%%
%%ALIAS=GrPo=hep-th/0009211,hep-th/0106014%%
%%ALIAS=KST=hep-th/0201028,hep-th/0211182%%
%%ALIAS=BeBe=hep-th/0301161,hep-th/0310058%%
%%ALIAS=sugraIIB=Schwarz:1983wa,Schwarz:1983qr%%
%%ALIAS=BoCa=Bodner:1989cg,Bodner:1990zm%%
%%ALIAS=BeCv=hep-th/0403049,hep-th/0407263%%
%%ALIAS=BeJe=hep-th/0302047,hep-th/0311119%%
%%ALIAS=Gaunt=hep-th/0205050,hep-th/0212008,hep-th/0302158%%
%%ALIAS=Paris=hep-th/0206213,hep-th/0303127,hep-th/0406137,hep-th/0412187%%
%%ALIAS=mtensor=hep-th/0312210,hep-th/0407205,hep-th/0410051%%
%%ALIAS=instanton=hep-th/0208145,hep-th/0404147%%

%%%%%%%%%%%%%%%%%%%%%%%%%%%%%%%%%%%%%%%%%%%%
%% MAINMATTER
%%%%%%%%%%%%%%%%%%%%%%%%%%%%%%%%%%%%%%%%%%%%

\section{Introduction}

%%%%%%%%%%%%%%%%%%%%%%%%%%%%%%%%%%%%%%%%%%%%%%%%%%

One of the major problem appearing in compactifications of string
theory, is the emergence of a continuous moduli space of string
vacua. In order to get contact, not only to the standard model of
particle physics, but also to (inflationary) cosmology, these moduli
have to be fixed and if supersymmetry is broken only at fairly low
energies, we have to understand the mechanism while preserving at
least some supersymmetries. Moduli appear in two guises: the closed
string or geometrical moduli, related to deformations of the size and
shape of cycles of the internal manifold and open string moduli,
related to un-fixed positions of wrapped branes.  Fluxes provide a
mechanism to lift both moduli. On one hand, fluxes cause a
gravitational force which expands/contract a cycle that is
parallel/perpendicular to the flux and these competing effects yield
the stabilization at values related to the strength to the different
fluxes. Note, if a given cycle is parallel or perpendicular to all
fluxes, it is not stabilized and the supergravity potential has a
run-away mode. On the other hand, fluxes couple also to the world
volume action of branes producing a potential for the open string
moduli. Since the open string moduli are compact, any potential has an
extremum and gives a stabilization of these moduli.

Typically, one refers to fluxes as non-zero expectation values of the
RR- and NS-fields in the vacuum and there is a growing literature on
compactifications in the presence of fluxes
\cite{fluxold} -- \cite{hep-th/0502200}.
Fluxes are sourced by branes which give rise to $\delta$-function
singularities in the equations of motion or Bianchi identities and in
typical flux compactifications one does not take into account source
terms. But note, the constraints coming from solving the Killing
spinor equations are local equations and are not sensible to
distinguish between fluxes sources by branes and background fluxes.
In fact, often one can regard the background fluxes as a near horizon
limit for the corresponding branes. In addition to the fluxes related
to form-fields one can also consider metric fluxes (or both),
which are known as twisting. All fluxes can be related to a generalized
Scherk-Schwarz reduction, which one can apply to any global symmetry
of the vacuum. Generically, this reduction does not commute with all
supersymmetry transformations and hence supersymmetry is partially
broken. The original proposal of Scherk and Schwarz was related to a
global fermionic symmetry and breaks all supersymmetries
\cite{Scherk:1978ta}, but when applied to (global) gauge symmetries of
RR- and/or NS-forms, some supersymmetries remain unbroken.  In the
following we will only consider Scherk-Schwarz reductions that
correspond to flux compactifications and it is straightforward to
relate these reductions to gauged supergravity; see
\cite{hep-th/9601150} -- \cite{hep-th/0503114}
for more details.  This gives us a powerful tool, because in gauged
supergravity we can calculate the potential explicitly and expand the
fields around an extremum; although it may become technically very
involved. We should also note, that Scherk-Schwarz reductions allow
for a consistent truncation to the massless Kaluza-Klein spectrum.

Without relying on a specific Scherk-Schwarz reduction, the moduli
problem can also be addressed in gauged supergravity directly, e.g.\
by asking the question: Which gauging yields a potential that fixes
all moduli? In answering this question, one has to keep in mind that
massive tensor or vector multiplets imply further corrections to the
supergravity Lagrangian \cite{mtensor}. As a spin-off, one can also
look for possible deSitter vacua of the explicitly known potential in
gauged supergravity.  Unfortunately, in this approach the lift to
10-dimensional string theory, or M-theory, remains obscure in many
interesting examples.  Also the (deformed) geometry of the internal
space is only for very specific situations known and to address this
question, one has to solve directly the 10-dimensional (Killing
spinor) equations. The fluxes are then identified with specific
torsion components that deform the internal geometry. Unfortunately,
the reduction to 4 dimensions as well as the derivation of the Kahler
and superpotential as function of the scalar fields becomes a
non-trivial task in this approach.

Let us add another important remark. Having fixed the moduli may not
be enough. Especially interesting are vacua that remain stable if
supersymmetry is broken. Many vacua, obtained in gauged supergravity,
have unfortunately some tachyonic directions and if the cosmological
constant is lifted to a positive value, these models are not
acceptable. Of course, this is a problem of how we break supersymmetry
and what are the corrections to the potential.

We have organized this paper as follows.  We start in Section 2 with a
resume of the Kaluza-Klein reduction yielding a continuous (moduli)
space of vacua.  In Section 3 we explain the relation between
Scherk-Schwarz reductions and gauged supergravity, which provides a
tool to understand the lifting of the moduli space.  Section 4
is devoted to the second approach, ie.\ we solve the 10-dimensional Killing
spinor equations and relate fluxes to torsion components and
$G$-structures. Two examples are given, where the back reaction of the
fluxes changes the internal geometry and which are worked out
explicitly.

%%%%%%%%%%%%%%%%%%%%%%%%%%%%%%%%%%%%%%%%%%%%%%%%%%%%%%%%%%%

\section{Moduli coming from Kaluza-Klein reduction}

%%%%%%%%%%%%%%%%%%%%%%%%%%%%%%%%%%%%%%%%%%%%%%%%%%%%%%%

We start with a short summary of the standard Kaluza-Klein (KK)
reduction of type II supergravity ({\em without} fluxes). The reader
may consult the original papers \cite{BoCa} for further details.

In the low energy approximation, type II string theory is described by
type II supergravity in 10 dimensions, which has \N=2 supersymmetry
and can be chiral (type IIB) or non-chiral (type IIA) if the two
gravitinos and dilatinos in the fermionic sector have equal or
opposite chirality. Both supergravities have a common sector
comprising the (NS-NS) fields
\[
(e^m \ , \ B \ , \ \phi ) 
\]
where $e^m$ is the vielbein 1-form, $B$ is the NS-2-form and $\phi$
denotes the dilaton. But both models differ in the RR-sector.  On the
IIA side, there are odd RR gauge potentials
\be010
{\rm type\ IIA}:  \qquad (C_1 \ , \ C_3 ) 
\ee
which give rise to the following gauge invariant field strengths
\be011
F^{(2)} = m B + dC_1 \quad , \qquad
F^{(4)}= dC_3 + \frac{6}{m} F^{(2)} \wedge F^{(2)}\ ,
\ee
satisfying the Bianchi identities
\be620
dF^{(2)} = m H  \quad , \qquad  dF^{(4)} = 12 H \wedge F^{(2)} \ .
\ee
We included here also the mass parameter $m$, which was
introduced in supergravity by Romans \cite{Romans:1985tz} and is
related in string theory to D8-branes (``at the end of the
universe''). If $m \neq 0$, the RR-1-form $C_1$ can be gauged away
giving a mass to the NS 2-form $B$.  On the other hand, type IIB
supergravity has even RR gauge potentials
\be030
{\rm type\ IIB}: \qquad (C_0 \ , \ C_2 \ , \ C_4^+) 
\ee
and one defines the following field strengths
\be031 
P = \frac{1}{1- |T|^2} dT \ , \quad
G^{(3)} = \frac{1}{
\sqrt{1- |T|^2}} (F_{3}- T F_{3}^\star) \ , \quad 
F^{(5)}= dC_{4}-\frac{1}{4}\,( C_{2}\wedge dB) 
\ee
where the 5-form $F^{(5)}$ has to satisfy a self-duality constraint and
\be032
F_{3} = d(B+iC_{2}) \quad  ,
\qquad T = \frac{1+i\tau}{1-i\tau} \quad  , \qquad   \tau = C_0 + i e^{-\phi}
\  .
\ee
The Bianchi identities read now
\be772
dG^{(3)} = (i Q -P) \wedge G \quad , \qquad
dF^{(5)} = - \frac{1}{4} dC_2 \wedge dB \ .
\ee
with the U(1) connection
\[
Q= \frac{1}{1- |T|^2} {\rm Im}(T d T^\star) \ .
\]
Type IIB supergravity has an SL(2,R) symmetry, which acts as: $\tau
\rightarrow \frac{a \tau + b }{ c \tau +d}$ combined by a rotation of
the 2-form doublet with the matrix: $\Big(\! \ba{cc} a&b \\[-1mm]
c&d\ea \!\Big)\in SL(2,R)$. This symmetry can also be used to write
the 3-form as: $G_3 \sim i \, e^{\phi/2} (F_3 + \tau \, H_3)$, which
is more standard in type IIB string theory [note, we use the Einstein
frame, which explains the factor $e^{\phi/2}$].

In the KK reduction one splits first the 10-dimensional space into the
4-dimensional external space ${\cal X}$ and the internal space ${\cal
Y}$ with the coordinates $x^M = \{ x^\mu , y^m\}$ ($\mu, \nu , ... =
0\ldots 3$ , $m = 4\ldots 10$), followed by an integration over the
internal coordinates. This is straightforward if the fields do not
dependent on the internal coordinates $y^m$. In general however, this
is not the case and one has to make a Fourier expansion of the fields
in a complete set of harmonic eigenfunctions on the internal
manifold. {From} the 4-dimensional point of view the higher Fourier
modes correspond to massive excitations and only these excitations
depend on the internal coordinates.  The assumption that the fields do
not depend on the internal coordinates is therefore equivalent to a
truncation of the KK-spectrum on the massless sector. This is
consistent only if the massless KK-fields do not act as sources for
massive KK-fields, which is ensured if the Fourier expansion of the
Lagrangian does not contain a coupling that is {\em linear} in the
massive fields. Otherwise, the equations of motion are not solved by
simply setting the massive modes to zero.  This question can be
answered if the internal metric is explicitly known -- although the
concrete calculation might be very involved.  It has been done so far
only for very few examples as e.g.\ for the sphere compactification of
11-dimensional supergravity with 4-form \cite{deWit:1986iy}. If the
internal metric is not known, which unfortunately is often the case,
the consistent truncation on the massless sector becomes a highly
non-trivial problem. But one can formulate the truncation nevertheless
in a weaker sense, where one admits the disturbing couplings so that
massless fields appear as sources for massive fields, but only via
higher derivatives.  In this case, one cannot simply set to zero the
massive fields, but has to integrate them out. Due to the derivative
coupling, this does {\em not} change the effective low energy
Lagrangian up the two derivative level but modifies higher derivatives
terms (which have been neglected anyway in the 10-dimensional
Lagrangian). This procedure of integrating out of massive modes is
only consistent if it can be done in a finite number of steps, ie.\ if
there are sources for only finite number of massive fields. If this is
not the case, the theory remains genuinely higher dimensional and one
cannot perform a KK reduction. We will assume that this is not the
case for our models. For details we refer to \cite{Duff:1989cr} where
the analysis for a Calabi-Yau space is given.

If there are {\em no} fluxes, the internal space has to be Ricci-flat
and supersymmetry requires that the space has to have restricted
holonomy.  For a 6-dimensional internal space this means that the
holonomy can be at most SU(3), ie.\ strictly inside SO(6) and the
space is called Calabi-Yau. The amount of supersymmetry in four
dimensions depends on the number of Killing spinors on the internal
space and because the Killing spinors have to be singlets, this number
is directly related to the holonomy. For SU(3) it is exactly one
internal spinor and therefore 1/4 of supersymmetry is broken; note a
6-dimensional flat space allows four independent Weyl spinors [transforming
under the ${\bf 4}$ of $SU(4) \simeq SO(6)$]. After the reduction, the
resulting 4-dimensional theory will have eight supercharges or has
\N=2, D=4 supersymmetry and therefore supergravity can couple to
vector and hyper multiplets; in ungauged supergravity tensor
multiplets are dual to hyper multiplets.  Each multiplet has four
bosonic degrees of freedom: the graviton and graviphoton in the
gravity multiplet; a vector field and one complex scalar in the vector
multiplets and each hyper multiplet contains four real scalar fields.

The moduli are the zero modes of the scalar fields and appear in KK
reduction from two sources: from the internal metric components and
from the RR/NS-forms. They are in one-to-one correspondence to
harmonic forms on the internal space which, for a 6-dimensional space
with SU(3) holonomy, are equivalent to deformations of the complex
structure and Kahler class as well as to the gauge symmetries of the
RR/NS-form potentials. For example, an n-form $\omega_n \neq
d\omega_{n-1}$ on the internal space with: $d\omega_n = d
{^\star\omega_n}=0$ gives a scalar field $\phi=\phi(x)$ in the Fourier
expansion $C_n = \phi(x) \, \omega_n + \ldots $, where $C_n$ is an
n-form potential. The corresponding field strength is given by $F_n =
d\phi \wedge \omega_n$ and therefore a modulus in the low energy
theory is given by the constant part of $\phi$.  The appearance of
this modulus is of course a consequence of the gauge symmetry for the
form field.  The moduli coming from the internal metric are related to
deformations of the Kahler 2-form $J$ and the holomorphic 3-form
$\Omega$.

To be concrete the expansion goes as follows; see \cite{BoCa}
for more details. We denote a complete basis of harmonic 3-forms
by\footnote{As usual we distinguish between imaginary selfdual and
anti-selfdual forms, which are complex conjugate to each other.}:
$\{\chi^k , \tilde \chi_k \} \in H^{(3)}({\cal Y})$ and the basis of
harmonic 2-forms by: $\{\omega_a\} \in H^{(2)}({\cal Y})$, which 
by Hodge duality are related to 4-forms spanning $H^{(4)}({\cal Y})$.
Apart from the trivial 0-form and the volume form, there are no
(regular) 1- nor 5-forms on a Calabi-Yau manifold and we have to
expand all fields in these two sets of forms.

On the IIA side, we expand the RR-3-form and NS-2-form as well as the
Kahler class as follows
\be070
C_3 = u^k \chi_k + A^a \wedge \omega_a +  cc \quad , \qquad
B +iJ = z^a \omega_a 
\ee
where we denote the Kahler class with $J$ and $A^a$ are the
4-dimensional KK gauge 1-forms.  Since there are no harmonic 1-forms
on a Calabi-Yau space, $C_1$ does not give rise to a KK-scalar and
becomes the graviphoton upon dimensional reduction. The complex
scalars $u^k$ together with another set of complex scalars $v^k$
related to complex structure deformations enter hyper multiplets and
the complex scalars $z^a$ enter vector multiplets. One of the simplest
cases are the rigid Calabi-Yau spaces that do not have complex
structure deformations and hence none of the scalars $\{u^k , v^k\}$
are present. In this case, only one hyper multiplet is non-trivial and
this is the so-called universal hyper multiplet, which consists of the
dilaton $\phi$, the external $B$-field component (dualized to a
scalar) and the (3,0) and (0,3) part of the RR-3-form. Although there
are (rigid) Calabi Yau spaces without any harmonic (2,1)- and
(1,2)-forms (e.g.\ ${\mathbb T}^6/{\mathbb Z}_3$ has $h^{(2,1)}=0$),
we should stress that the (3,0)- and (0,3)-forms are always
non-zero. Explicit BPS solutions in 4 dimensions have a timelike or
null Killing vector and the stationary case has been discussed in
\cite{hep-th/9705169}.

On the IIB side the NS-2-form and RR-forms are decomposed as follows
\be080
C_4 = A^k \wedge \chi_k + cc  \quad , \qquad {^{\star_6} C_4} + i \, C_2
= v^a \omega_a 
\quad , \qquad B + i J = u^a \omega_a 
\ee
(${^{\star_6} C_4}$ denotes the 6-dimensional Hodge dual of the
4-form). Now, the four scalars of a hyper multiplet are given by the
two complex scalars: $\{u^a , v^a\}$. Since the Kahler class is always
non-trivial for a Calabi-Yau space, we obtain at least two hyper
multiplets.  One of them is the universal hyper multiplet which now
comprises the axion-dilaton $\tau$ combined with the (dualized)
external components of the NS- and RR-2-forms. On the IIB side the
scalars in the vector multiplets are related to deformations of the
complex structure, ie.\ come from the components of the internal
metric. For a rigid Calabi-Yau (ie.\ $h^{(2,1)}=0$) all vector
multiplets are trivial and, apart from the gravity multiplet, we have
only hyper multiplets (and the typical BPS solutions are the instanton
solutions as discussed in
\cite{hep-th/9706096,instanton}).

Since the complex structure and Kahler class deformations for a
given Calabi-Yau are not related to each other, the corresponding
moduli spaces appear as a direct product in the low energy
supergravity: {${\cal M} = {\cal M}_V \times {\cal M}_H$}, where the
vector multiplet moduli space ${\cal M}_V$ is a special Kahler
manifold and the scalars in hyper multiplets parameterize a
quaternionic Kahler space ${\cal M}_H$. Since these Kaluza-Klein
reductions do not give rise to a potential, the scalars can take any
constant value in the vacuum. In order to lift this moduli space
by giving a vev to the scalar fields, one has to generate a potential
upon compactification. This can be done by taking into account nonzero
fluxes.

%%%%%%%%%%%%%%%%%%%%%%%%%%%%%%%%%%%%%%%%%%%%%%%%

\section{Fluxes, Scherk-Schwarz reduction and gauged supergravity}

%%%%%%%%%%%%%%%%%%%%%%%%%%%%%%%%%%%%%%%%%%%%%

If we include fluxes, the dimensional reduction or compactification
becomes more involved. On the other hand since flux compactifications
are related to gauged supergravity not only the complete 4-dimensional
Lagrangian can be constructed but also the issue of moduli
stabilization can be addressed. We are now following in part the
literature as given in \cite{fluxold,hep-th/0202168}.

One may argue that in the vacuum all fields should be trivial and the
metric is (Ricci-) flat.  This strong restriction is not justified,
and non-zero values of RR- and NS-fields can still be considered as a
viable vacuum configuration -- at least as long as they respect the
{\em 4-dimensional} Poincar{\'e} symmetry. If so, we are dealing with
compactifications in presence of fluxes and one can distinguish
between gauge field and metric (or geometric) fluxes, that are related
by supersymmetry.  This means, that the form fields (\ref{010}) or
(\ref{030}) are non-trivial in the vacuum, but nevertheless obey the
equations of motion and Bianchi identities. These fluxes generate a
non-zero energy-momentum tensor and hence the internal metric is in
general not Ricci-flat; the resulting geometries can be quite
complicated (as we will see later).

To make this more explicit let us note, that in the simplest case
gauge field fluxes can be generated by a linear dependence (on the
internal coordinates) of the KK scalars coming from gauge fields in
(\ref{070}) or (\ref{080}).  The gauge symmetry implies that these
scalars appear only via derivatives in the Lagrangian/equations of
motion and hence, a linear dependence on the internal coordinates
leaves the Lagrangian still independent of the internal coordinates
and one can integrate over them. This is known as the (generalized)
Scherk-Schwarz reductions, which can be applied to any global symmetry
and especially to those KK scalars, that parameterize an isometry of
the moduli space. If the scalar field comes from an internal metric
component, the underlying global symmetry is related to specific
coordinate transformations and a linear dependence of these scalars is
also known as metric fluxes or twisting. In general, this procedure
does not commute all supersymmetry transformations and hence
supersymmetry is at least partially broken.  In the original
Scherk-Schwarz reduction this was done with respect to a fermionic
phase transformation, which did not commute with the supersymmetry
transformation and hence supersymmetry was broken completely
\cite{Scherk:1978ta}. In the case here, where we apply it to
isometries of the moduli spaces, some supersymmetries remain unbroken,
or in other words, some supersymmetry transformations commute with the
(generalized) Scherk-Schwarz reduction.  Nevertheless, masses for
scalars and vectors are generated and hence (part of) the moduli space
is lifted. More details on these reductions are given in the
literature, see \cite{hep-th/9601150} -- \cite{hep-th/0503114}.

{From} the lower dimensional point of view Scherk-Schwarz reductions
correspond to a gauging of the corresponding global symmetry and
following \cite{hep-th/9707130} let us discuss a simple example.  If
we just keep the axion-dilaton coupling, the type IIB supergravity
action reads
\be092
S \sim \int \sqrt{g} \, \Big[ R \, - \,
 \frac{g^{MN} \partial_M \tau \partial_N \bar \tau
}{ |\tau - \bar \tau|^2} \Big]
\ee
which exhibits, as part of the SL(2,R) symmetry, the axionic shift
symmetry
\[
\tau \rightarrow \tau + c
\]
for any $c=const$. In the Scherk-Schwarz reduction over one coordinate
(say $y$) one assumes $c = m y$ and hence one writes
\be091
\tau(x,y) = \tau(x) + m y \ .
\ee
For the metric, one makes the usual KK-Ansatz 
\[
ds^2 = e^{2\sigma} (dy + A_\mu dx^\mu)^2 + g_{\mu\nu} dx^\mu dx^\nu \\[2mm]
\]
where $\partial_y$ is a Killing vector. The inverse metric becomes
\[
g^{MN} \partial_M \partial_N \ =\ 
e^{-2\sigma} \partial_y \partial_y + g^{\mu\nu} {\cal D}_\mu {\cal D}_\nu
\]
with the covariant derivative
\[
{\cal D}_\mu = (\partial_\mu - A_\mu \partial_y) \ .
\]
Thus, the kinetic term yields
\[
\frac{g^{MN} \partial_M \tau \partial_N \bar \tau
}{ |\tau - \bar \tau|^2} \ =
\  \frac{g^{\mu\nu} D_\mu \tau(x) D_\nu \bar \tau(x)
}{ |\tau(x) - \bar \tau(x)|^2} + \frac{ m^2 }{ |\tau(x) - \bar \tau(x)|^2}
 \; e^{-2 \sigma}
\]
where the second term is a (run-away) potential and the covariant
derivative in the kinetic term is
\be821
D_\mu \tau =\partial_\mu \tau(x) - m A_\mu \ .
\ee
Therefore, the scalar field ${\rm Re} (\tau)$ is now charged with
respect to the {\em local} shift transformations
\[
\tau \rightarrow \tau + c(x) \quad , \qquad  
A \rightarrow A + \frac{1}{ m} dc 
\]
and in the original metric the gauge transformation in $A$ can be
absorbed by a coordinate transformation $y \rightarrow y + c(x)$.
Obviously, the same result can be obtained by a gauging of the global
shift symmetry in the reduced theory.  The charged scalar, ${\rm
Re}(\tau)$, does not enter the potential and represents a flat
direction, which is required by gauge invariance and which in turn can
be used to gauge away the scalar giving a mass to the gauge boson (the
kinetic term for ${\rm Re}(\tau)$ becomes a mass term for $A_\mu$).
As we will see next there is also the dual situation where not a
vector becomes massive, but an antisymmetric tensor becomes massive by
``eating'' a vector\footnote{Note, in 4 dimensions a {\em massive}
vector is dual to a {\em massive} tensor.}. But before we come to
this, let us note, that the run-away behavior of the potential, ie.\
the absence of a fixed point, is related to the non-compactness of the
gauged isometry and does not happen if the isometry is a U(1) action,
that is not freely acting \cite{hep-th/0104056}.

This Scherk-Schwarz reduction was related to the isometry $\partial_y$
and generated an internal flux given by the 1-form: $d_y \tau = m dy$
and as result the scalar field ${\rm Re}(\tau)$ became charged under
the corresponding KK vector field. A general Calabi-Yau space has no
isometries and the internal metric does not give rise to 4-dimensional
vector fields, but nevertheless there is an analogous mechanism which
relates flux compactification to gauged supergravity.  To be concrete
we follow now \cite{hep-th/0012213,hep-th/0202168}, consider the type
IIB case with fluxes for the NS-2-form $B$ and write for (\ref{080})
\[
B +i J = u^a(x,y) \omega_a = [u^a(x)\,  + c^a(y) ]\,  \omega_a \ .
\]
The coefficients $c^a(y)$ are fixed by the requirement that the
corresponding field strength yields a real {\em internal} 3-form
(=flux), which can be expanded in the basis $\{\chi^k, \chi_k\}$ with
the {\em constant} coefficients $m_k$, ie.\ $d c^a(y)\wedge \omega^a =
m^k \chi_k = H^{flux}$ giving
\[
d(B+i J) = du^a(x)\,  \wedge \omega_a + (m^k \, \chi_k + cc) \ .
\]
To keep the notation simple, we drop all indices ($m^k \rightarrow m$)
and collecting the terms containing this mass deformation yields for
the 5-form
\[
F_5 \ = \ dC_4 - \frac{1}{ 4} C_2 \wedge H \ =\ 
[ dA -  \frac{1}{ 4} m \, C_2^{(ext)} ] \wedge
\chi + cc + \cdots 
\]
where $C^{ext}_2$ is the external component of the 2-form [note, a
Calabi-Yau has no non-trivial 5-forms and therefore $\omega \wedge
\chi =0$]. Now, the kinetic term for this 5-form yields exactly a
massive 2-form coupling in 4 dimensions
\[
\Big[(dA)_{\mu\nu} -  \frac{1}{8} \, m \, C_{\mu\nu}^{(ext)} \Big]^2 
\]
and this expression has to be dressed up with the metric of the
complex structure moduli space (we suppressed the indices). This
massive 2-form can be dualized to a massive vector, where the charged
scalar is given by the dual of $C_2^{(ext)}$ and which enters the
universal hyper multiplet.  The potential comes now from the 3-form:
$H_3= du \wedge \omega + m \chi + cc$, which yields after squaring a
term: $|m|^2 \equiv G^{ab} m_a m_b$ (actually it might be useful to
introduce here symplectic notation).  It is an important property of
type IIB compactifications in the presence of 3-form fluxes, that the
potential is always positive definite (no-scale form) and therefore
the only supersymmetric vacua are flat space vacua. This property is
however lost for more general flux compactifications.

It is now straightforward to consider fluxes also for the 2-form
$C_2$, which makes the external components of the NS $B$-field
massive. In general, one can also consider a combination of both
3-form fluxes, but due to the Chern-Simons terms, we cannot consider
independent massive deformations with respect to both 2-forms. This is
also reflected by the fact, that both shifts do not correspond to two
commuting isometries on the quaternionic space parameterized by the
universal hyper multiplet.

One should have expected that each flux compactification is related to
a specific vacuum of gauged supergravity, although the concrete
embedding might be involved. The opposite statement, ie.\ whether
every vacuum obtained in gauged supergravity can be embedded into a
specific flux compactification, is far from clear -- most likely this
is not possible. We have to keep in mind that one can also consider
metric fluxes, which are related to Scherk-Schwarz reductions with
respect to axionic scalars of the internal metric, which we did not
discussed here, see \cite{hep-th/9901045,hep-th/0210209}. This becomes very
involved, if one wants to do it explicitly and it might not be
necessary because the realization of flux compactifications within
gauged supergravity opens the possibility to understand the
moduli stabilization within gauged supergravity and we shall summarize in
the following some essentials.

The starting point is the Lagrangian of ungauged supergravity with
\N=2 supersymmetry in 4 dimensions, which is obtained from standard
Kaluza-Klein reduction as discussed in the previous section giving
rise to a continuous moduli space: ${\cal M} = {\cal M}_V \times {\cal
M}_H$, where ${\cal M}_{V}$ and ${\cal M}_{H}$ are parameterized by
the scalars belonging to the vector multiplets and to the hyper
multiplets, respectively. Potentials that are consistent with ${\cal
N}=2$ supersymmetry are obtained by performing a gauging of various
global symmetries.  There are two different types of gaugings, namely
$(i)$ one can either gauge some of isometries of the moduli space of
ungauged ${\cal N}=2$ supergravity or $(ii)$ one can gauge (part of)
the $SU(2)$ R-symmetry, which only acts on the fermions.  We are
interested in a gauging that generate a potential for both types of
scalars, because we want to derive constraints for lifting the
complete moduli space. In the following we will therefore discuss
gaugings of isometries of ${\cal M}_H$ and refer to
\cite{hep-th/9605032} for a detailed description of ${\cal N}=2, D=4$
gauged supergravity.

Scalar fields in hyper multiplets parameterize a quaternionic Kahler
manifold ${\cal M}_{H}$ and these spaces possess three complex
structures $J^x$ as well as a triplet of Kahler two-forms $K^x$ ($x
=1,2,3$ denotes the $SU(2)$ index).  The holonomy group of these spaces 
is $SU(2) \times Sp(n_H)$ and the Kahler forms are covariantly
constant with respect to the $SU(2)$ connection. The isometries of
${\cal M}_{H}$ are generated by a set of Killing vectors $k_I = k_I^u
\partial_u$
\be090
q^u \rightarrow q^u + k^u_I \epsilon^I 
\ee
where ``$I$'' counts the different isometries and $q^u$ are the scalar
fields of hyper multiplets. The gauging of (some of) the Abelian
isometries gives gauge covariant derivatives $dq^u \rightarrow d q^u +
k_I^u A^I$ so that the vector field become massive.
In order to maintain supersymmetry, the gauging has to
preserve the quaternionic structure, which implies that the Killing
vectors have to be tri-holomorphic, which is the case whenever it is
possible to express the Killing vectors in terms of a triplet of real
Killing prepotentials ${\cal P}^x_I$ as follows:
\be081
K^x_{uv} k^v_I = - \nabla_u {\cal P}_I^x \equiv - \partial_u {\cal P}_I^x -
\epsilon^{xyz} \omega^y_u {\cal P}^z_I 
\ee
where $\omega_u^y$ is the $SU(2)$ connection giving the Kahler forms
by $K^x_{uv} = - \nabla_{[u} \omega^x_{v]}$.  By using the Pauli
matrices $\sigma^x$ one can also use a matrix notation: ${\cal P}_I =
\sum_{x=1}^3 {\cal P}^x_I \, \sigma^x$. With these Killing
prepotentials one can define an SU(2)-valued superpotential by
\cite{hep-th/0101119,hep-th/0104056}
\be927
W^x = X^I {\cal P}_I^x \equiv X^I(z) {\cal P}^x_I(q) \ ,
\ee
where $\{z , q \}$ denote collectively the scalars from vector and
hyper multiplets. A real valued superpotential can be defined as $W^2
= e^{K} \det (W^x \sigma^x)$, where $K$ is the Kahler potential of the
special Kahler manifold ${\cal M}_V$. Supersymmetric vacua are extrema
of the real superpotential, which are equivalent to a covariantly
constant superpotential $W^x$. This gives as constraints
for supersymmetric vacua
\be621
(i) \quad (\nabla_A X^I) {\cal P}_I^x = 0 \quad , \qquad (ii)
\quad X^I (\nabla_u
{\cal P}_I^x) = K^x_{uv} (X^I k_I^v) = 0 
\ee
where $\{\nabla_A, \nabla_u\}$ denote the Kahler/SU(2)-covariant
derivatives with respect to the scalars $\{z^A , q^u\}$ in
vector/hyper multiplets and $X^I = X^I(z)$ is part of the symplectic
section $(X^I, F_I)$ [$F_I$ is the derivative of the prepotential
$F(X)$ with respect to $X^I$].

For concrete models one has to know the moduli spaces, which are are
more or less well known in the classical limit and many gaugings have
been considered already. On the vector multiplet side, all corrections
(perturbative and non-perturbative) are included in a prepotential,
which on the 2-derivative level is holomorphic and homogeneous of
degree two.  Much less is known about quantum corrections for the
hyper multiplet moduli space; the 1-loop correction has recently been
found in \cite{hep-th/0307268} and instanton corrections are discussed
in \cite{instanton}. Classically, it is given by
the coset space $SU(2,1)/U(2)$, which is one of the few spaces that
are quaternionic and Kahler at the same time and its Kahler potential
can be written as
\be140 
K = - \log [S+\bar S - 2(C + \bar C)^2] \;.
\ee 
This coset space has two commuting Abelian isometries which are
generated by the Killing vectors associated to shifts in the imaginary
parts of $S$ and $C$ and their gauging has been discussed in
\cite{fluxold,hep-th/0101119,hep-th/0104056,hep-th/0101007}. 
The gaugings of these two shift symmetries correspond exactly to the
Scherk-Schwarz reductions that we discussed at the beginning of this
section (the $S$ shift corresponds to the $\tau$ shift and the
$C$-shift to the 3-form flux).

%%%%%%%%%%%%%%%%%%%%%%%%%%%%%%%%%%%%%%%%%%%%%%%%%%%

\bigskip

\qquad {\bf Can one fix all moduli in gauged supergravity?}

\medskip

%%%%%%%%%%%%%%%%%%%%%%%%%%%%%%%%%%%%%%%%%%%%%%%%%%%

\noindent
In order to fix the moduli from the vector as well as hyper multiplet
it was important that we gauged an isometry of the quaternionic space
${\cal M}_H$ or equivalently to add fluxes (3-form flux on the IIB
e.g.) which make a scalar of a hyper multiplet massive. This is only
the minimal requirement, on top of this gauging one may also consider
to gauge isometries of vector multiplet moduli space ${\cal M}_V$. The
resulting superpotential obtained from gauged quaternionic isometries
was given in (\ref{927}) with $X^I = X^I(z)$ as the ``electric'' part
of the symplectic section $V =(X^I , F_I)$.  It is a known problem,
that gauged supergravity prefers the electric part and does not
produces the magnetic part of the superpotential.  But by taking into
account also (massive) tensor multiplets, one can promote it to a
manifestly symplectic expression \cite{mtensor}. An important property
of this setup is however, that by a symplectic transformation one call
always go into a strictly perturbative regime where all magnetic
charges vanish so that the potential in (\ref{927}) can always be
considered. This property that the electric and magnetic charges are
mutually local is a consequence of supersymmetric Ward identities
\cite{mtensor}.

We can now discuss the conditions of getting a complete lifting of the
moduli space.  A necessary condition for this is that the variations
of the hyperino and gaugino vanish for constant scalars which yielded
eqs.\ (\ref{621}). The condition $(ii)$ is equivalent to the
existence of a fixed point for the Killing vector
\[
k=X^I k_I
\]
and the {\em complete} hyper multiplet moduli space is lifted if $k$
has a NUT fixed point, ie.\ if it represents a {\em point} on ${\cal
M}_H$.  This excludes by the way, axionic shift symmetries and
requires a {\em compact} isometry \cite{hep-th/0104056}.  The fixed
point set of a Killing vector field is always of even co-dimension,
which is related to the rank of the 2-form $dk$ calculated on the
fixed point set.  For a related discussion see \cite{999}. In fact, if
the rank is maximal, ie.\ $\det (dk) \neq 0 $, the fixed point set is
in fact a point on the manifold and $dk$ parameterizes a rotation
around the fixed point.  Otherwise, any zero mode of $dk$ would
parameterize a shift symmetry of the fixed point set and hence if
$\det (dk) =0$, the potential will have some flat directions.
Therefore, we get the following two conditions for lifting the hyper
multiplet moduli space
\be922
|k|=0 \quad , \qquad {\rm with:} \quad \det (dk) \neq 0 \ .
\ee
If we can find a Killing vector that satisfies both conditions, the
hyper multiplet moduli space will be lifted in the vacuum.  We should
place a warning here. Although, the isometries on the classical level
are well understood it is unclear whether the full quantum corrected
moduli space has isometries at all, which makes the moduli fixing
issue obscure -- at least from the supergravity point of view. But we
do not want to speculate here about the quantum moduli space for hyper
multiplets and shall instead continue with the discussion of the
second condition in (\ref{621}).  If the hyper scalars are fixed, the
Killing prepotentials are some fixed functions of the scalars of the
vector multiplets, ie.\ ${\cal P}_I = {\cal P}_I\big(q(z)\big)$ and
hence they vary over ${\cal M}_V$.  If ${\cal P}_I$ would be constant,
only one vacuum can occur, namely at the point where this constant
symplectic vector is a {\em normal vector} on ${\cal M}_V$
\cite{hep-th/0001082}. But since ${\cal P}_I$ varies now, it might
become normal at different points, related to the appearance of
multiple critical points as eg.\ the ones discussed in
\cite{hep-th/0104056}. If we calculate the second covariant
derivatives on ${\cal M}_V$ at this fixed point, ie.\ $\nabla_{\bar A}
\nabla_B X^I {\cal P}_I^x $ and use relations from special
geometry\footnote{Because: $\nabla_{\bar A} \nabla_B X^I \sim g_{\bar
A B} X^I$.} we find that all these critical points are isolated -- at
least as long as the metric does not degenerate. Therefore, there are
no further constraints from the vector multiplet moduli space and the
crucial relations that have to be realized are the ones in
(\ref{922}).

How about supersymmetric flat space vacua with all moduli
fixed?\,\footnote{I am grateful to Gianguido Dall'Agata for a discussion
on this point.} If the matrix $W^x \sigma^x$ does not degenerate, this
is only possible if we impose the additional requirement that $W^x$
vanish in the vacuum and this implies that the Killing prepotentials
have to vanish, ie.\ $0={\cal P}^x \equiv K^x_{uv} \partial^u k^v$. We
are interested in the case where all moduli are fixed and hence
$dk|_{|k|=0}$ has to be an orthogonal matrix. Recall, the holonomy of
a quaternionic space was $SU(2) \times Sp(n_H)$ and since the Kahler
forms $K^x$ are $Sp(n_H)$ singlets, the Killing prepotentials vanish
if the rotation parameterized by $dk|_{|k|=0}$ is in a subgroup of
$Sp(n_H)$ and leaves the $SU(2)$ part invariant. We leave it open,
whether there exist an appropriate Killing vector $k$ obeying all
constraints.

In the discussion so far we did not mention the fact that the
potential obtained in gauged supergravity is always independent of the
charged scalar field(s) and we have to ask whether this indicates some
flat directions of the potential.  There are two reasons why this does
not spoil our discussion so far.  On one hand, if the fixed point set
is zero-dimensional (ie.\ a point) this flat direction is only an
artificial angular coordinate on the moduli space. On the other hand,
as we mentioned already before, the charged scalars ``can be eaten''
by the vector fields giving them a mass that corresponds to the
eigenvalues of $(h_{uv}k^u_I k^v_J)|_{|k|=0}$. So, there are no moduli
related to these scalars anymore.

In gauged supergravity the problem of fixing the moduli is rather well
formulated. The question is however, to embed these models into string
or M-theory, ie.\ to obtain the potential by a suitable dimensional
reduction.  In addition, although these reductions may lift the moduli
space, it is not granted that we will obtain a unique vacuum and we
can end up with a landscape of string vacua \cite{hep-th/0307049}.
Although this conclusion was reached only for specific fluxes and it
is not inevitable for (most) general fluxes, it may happen that we
have to rely, at least to a certain extend, on an anthropic selection
for choosing the vacuum in which we live \cite{hep-th/0302219}.
Adopting this philosophy, the problem with the cosmological constant
disappears.  In gauged supergravity this ambiguity is related to the
different ways of gauging a given isometry with different U(1) gauge
fields.  On the other hand, if one takes into account general fluxes
on the type IIA side, the vacuum is rather unique \cite{BeCv} and one
may wonder whether the landscape is an ``artifact'' of Calabi-Yau
compactifications of string theory. This brings us to another approach
to address the consequences of fluxes and which allows directly to
determine the resulting deformed geometry.

%%%%%%%%%%%%%%%%%%%%%%%%%%%%%%%%%%%%%%%%%%%%%%%%

\section{Deformed geometry and $G$-structures}

%%%%%%%%%%%%%%%%%%%%%%%%%%%%%%%%%%%%%%%%%%%%%%%%

The approach to flux compactifications via gauged supergravity had the
advantage to yield the explicit potential and one can therefore not
only investigate supersymmetric vacua, but explore also possible
deSitter vacua. The drawback is however that this approach does not
yield the 10-dimensional geometry and to obtain this flux-deformed
geometry one has to solve directly the 10-dimensional supersymmetry
constraints.

Supersymmetry exchanges fermionic with bosonic degrees of freedom and
in a supersymmetric vacua with trivial fermions, the fermionic
variations have to vanish [the variations of the bosonic field vanish
identical for trivial fermionic fields].  For type II supergravity,
these are the gravitino $\delta\Psi_{\mu}$ (spin 3/2) and dilatino
$\delta\lambda$ (spin 1/2) variation. On the IIA side we can combine
both Majorana-Weyl spinors of opposite chirality to a general Majorana
spinor and the variations read in the string frame
\cite{hep-th/0103233}
\be040
\ba{rcl}
{\rm IIA:} \qquad 
\delta \psi_M &=& \Big\{D_M  + \frac{1}{ 8} H_M \Gamma_{11}
+ \frac{1}{ 8} \, e^{\phi} \Big[ \, m  \,  \Gamma_M  +
  F^{(2)} \Gamma_{M}  \, \Gamma_{11}
+  \, F^{(4)} \, \Gamma_M  \Big] \Big\} \epsilon \ ,
\\[3mm]
\delta \lambda &=&  \Big\{ \partial \phi 
      + \frac{1}{ 12} \, H \, \Gamma_{11} 
       + \frac{1}{4} e^{\phi}\Big[ 5\, m + 3 F^{(2)} \, \Gamma_{11}
      +  F^{(4)} \Big] \Big\}  \epsilon 
\ea
\ee
where $\epsilon$ is the Killing spinor which is also Majorana.  On the
IIB side, both Majorana-Weyl spinors have the same chirality and can be
combined into a single (complex) Weyl spinor so that the variations can be
written in the Einstein frame\footnote{Which is more appropriate on
the IIB side, because it makes the SL(2,R) symmetry manifest.} as
\cite{sugraIIB,hep-th/9805131}
\be006
\ba{rcl}
{\rm IIB:} \qquad 
\delta\Psi_{M} &=& \Big[D_{M} -\frac{i}{2} Q_M
+ \frac{i}{480} F^{(5)}\Gamma_{M} \Big] \epsilon 
-\frac{1}{96}\Big[G^{(3)} \, \Gamma_M + 6 G^{(3)}_{M} \Big] 
\epsilon^{\star}\ , \\[3mm]
\delta\lambda &=& i\, P \epsilon^\star-
\frac{i}{24} G^{(3)} \epsilon
\ea
\ee
where all fields were introduced in Section 2. In these variations all
indices are contracted with $\Gamma$-matrices, ie.\ we used the
abbreviations
\be050
\partial \equiv \Gamma^M \partial_M \  , \quad
F^{(2)} = F^{(2)}_{MN} \Gamma^{MN} \ , \quad 
H = H_{PQR} \Gamma^{PQR} \  , \   H_M = H_{MPQ} \Gamma^{PQ} \ ,
\quad {\rm etc .} 
\ee
[see also (\ref{011}) and (\ref{031}) for the definition of the
fields].

%%%%%%%%%%%%%%%%%%%%%%%%%%%%%%%%%%%%%%%%%%%%%%%

\bigskip

\qquad {\bf Killing spinors and $G$-structures}

\medskip
%%%%%%%%%%%%%%%%%%%%%%%%%%%%%%%%%%%%%%%%%%%

\noindent
The number of unbroken supersymmetries is given by the zero modes of
these equations, i.e.\ the number of Killing spinors for which these
variations vanish.  The 10-dimensional spinor can be expanded in all
independent internal and external spinors so that we can always write
\be542
\epsilon = \theta_l \otimes \eta_l 
\ee
with $\theta_l$ and $\eta_l$ denoting the four- and six-dimensional
spinors, respectively. On the IIA side, $\epsilon$ has to be Majorana
whereas on the IIB side it is Weyl.  In order to have a well-defined
KK-reduction, the internal spinors have to be singlets under the
structure group $G \subset SO(6)$. The internal space can admit up to
four Weyl spinors which transform under the ${\bf 4}$ of $SU(4) \simeq
SO(6)$ and therefore are singlets only under the trivial identity
(ie.\ structure group has to be trivial).  On the other hand, no
singlet spinors are possible if the structure group is the whole
SU(4); for SU(3) we find one singlet spinor, which appears as the
singlet in the SU(3) decomposition: ${\bf 4} \rightarrow {\bf 1 } +
{\bf 3}$. In the same way, if we have two independent internal
spinors, one can find an $SU(2) \subset SU(3)$ under which both are
singlets. If we do not take into account brane sources, the Killing
spinors have to be globally well-defined and the existence of these
spinors is in one-to-one correspondence to the existence of globally
well-defined differential forms. These differential forms which are
also singlet under the structure group $G$, define $G$-structures and
can be written as fermionic bi-linears
\be723
\Lambda^{(n)}_{kl} =  \eta^\dagger_k \gamma^{(n)} \eta_l
\quad , \qquad 
\Sigma^{(n)}_{kl} =  \eta_k^T \gamma^{(n)} \eta_l
\ee
where $\gamma^{(n)} \equiv \gamma_{m_1 m_2 \cdots m_n}$. By using the
Killing spinor equations one derives differential equations for the
internal spinors which in turn give differential equations for these
forms, see \cite{Gaunt}.  If we are interested in a 4-dimensional flat
vacuum, the external spinor is covariantly constant. The internal
spinor on the other hand, cannot be covariantly constant (due to the
fluxes) and solving the corresponding differential equation yields not
only the spinor but also the geometry of the internal space. Note, by
imposing the constraint that the spinor is a singlet under the group
$G$, the Killing spinor equations give a set of first order
differential equations for the vielbeine.

Having only one internal spinor, SU(3)-structures are given by
the 2-form and a 3-form
\be150
\eta^\dagger \, \gamma_{mn} \eta = i \, J_{mn}
\quad , \qquad
\eta^T \gamma_{mnp} \eta = i \, \Omega_{mnp} 
\ee
[with $1 = \eta^\dagger \eta$] where $J$ is a symplectic form with
$J^2 = - {\mathbb I}$ and can be used to define (anti) holomorphic
coordinates and $\Omega$ is then the holomorphic 3-form.  All other
fermionic bi-linear vanish as result of identities for 6-d
$\gamma$-matrices and hence no further (regular) singlet forms can be
build.  Being an SU(3) singlet spinor, $\eta$ satisfies the projectors
\be230
\ba{rcl}
(\gamma_m  + i J_{mn} \gamma^n ) \, \eta &=& 0 \ , \\[2mm]
(\gamma_{mn} -  i\, J_{mn}) \, \eta &=& \frac{i}{2} \,
\Omega_{mnp} \gamma^p  \, \eta^\star \ ,
 \\[2mm]
(\gamma_{mnp} - 3 i J_{[mn} \gamma_{p]} ) \, \eta &=& i \,
\Omega_{mnp} \eta^\star\ .
\ea
\ee
If the spinor is covariantly constant, these forms are closed and the
structure group is identical to the holonomy -- if not, the holonomy
is not inside SU(3) and the space cannot be Calabi-Yau (not even
complex in general). The failure of the structure group to be the
holonomy is measured by torsion classes.  Following the literature
\cite{630,Gaunt,hep-th/0211102,hep-th/0211118}, one
introduces five classes ${\cal W}_i$ by
\be526
\ba{rcl}
dJ &=& \frac{3 i}{ 4}\,  ( {\cal W}_1 \bar \Omega -
 \bar{\cal W}_1 \Omega ) +  {\cal W}_3 + J \wedge  {\cal W}_4 \ ,\\[2mm]
d\Omega &=&   {\cal W}_1 J \wedge J + J \wedge  {\cal W}_2
+ \Omega \wedge  {\cal W}_5
\ea
\ee
with the constraints: $J \wedge J \wedge {\cal W}_2 =J \wedge {\cal
W}_3 = \Omega \wedge {\cal W}_3=0 $.  Depending on which torsion
components ${\cal W}_i$ are non-zero, one can classify the geometry of
the internal space. E.g., if only ${\cal W}_1 \neq 0$ the space is
called nearly Kahler, for ${\cal W}_2 \neq 0$ almost Kahler, the space
is complex if ${\cal W}_1 ={\cal W}_2= 0$ and it is Kahler if only
${\cal W}_5 \neq 0$.

Before we will give examples, let us also comment on the SU(2) case,
we refer to \cite{hep-th/0403220,BeCv} for more
details.  In this case, we have two Weyl spinors $\eta^1$ and $\eta^2$
and we can define three 2-forms, that are supported on a 4-dimensions
subspace, and one holomorphic vector, which defines a fibration over
this 4-dimensional base space. The three 2-forms come on an equal
footing and one can pick one of them to use it as symplectic form and
the remaining two can be combined into one holomorphic 2-form so that
the 6-dimensional geometry is fixed by the triplet ($v, J^{(0)} , \hat
\Omega^{(2,0)}$).  If there are no fluxes, both spinors are
covariantly constant the internal space has SU(2) holonomy and is
therefore equivalent to ${\mathbb T}^2 \times K3$, where the three
(anti-selfdual) 2-forms are supported on K3 and the ${\mathbb T}^2$ is
identified by the vector field.  Naively, one would argue that SU(2)
structures can be relevant only for the very specific examples where
the Euler number of the 6-dimensional space vanishes; because only in
this case a globally well-defined vector field exist. But one has to
take the statement of ``globally well defined'' with a grain of salt,
because any wrapped brane can violate this requirement so that the
singular behavior is related to the location of the brane source. In
fact, instanton corrections coming from string world sheets and/or
instantonic 3-branes (which are wrapped on a 4-cycle) are typical
examples that imply SU(2) instead of SU(3) structure. Note, the
structures are related to certain fibrations of the manifold and have
no direct consequences of the amount of supersymmetry.

For the 4-dimensional external space, supersymmetry requires that it
has to be, up to warping, flat or anti-deSitter and hence we make the
Ansatz for the metric
\be060 
ds^2 = e^{2A(y)} \, \Big[ g_{\mu\nu} dx^\mu dx^\nu  + 
 h_{mn}(y) \, dy^m dy^n \, \Big] 
\ee
where $g_{\mu\nu}$ is either flat or $AdS_4$ and $h_{mn}$ is the
metric on ${\cal Y}$ and the warp factor depend only on the
coordinates of the internal space. In the vacuum all off-diagonal
terms should vanish and the fluxes should have only internal
components or are proportional to the 4-dimensional volume form. These
constraints are required by 4-dimensional Poincar{\' e} invariance.

Let us now discuss two flux vacua with SU(3) structure in more detail.

%%%%%%%%%%%%%%%%%%%%%%%%%%%%%%%%%%%%%%%%%%%%%%%%%%%%%%

\bigskip

\qquad {\bf Type IIA flux vacuum with SU(3) structure}

\medskip

%%%%%%%%%%%%%%%%%%%%%%%%%%%%%%%%%%%%%%%%%%%%%%%%%%

\noindent
On the type IIA side the list of literature about flux vacua is not as
long as on the IIB side, which is in part due to the stronger back
reaction of fluxes on the geometry and the difficulties in formulating
quantized string theory on these geometries. The internal geometry on
the IIB side on the other hand, remains complex which on the IIA side
this is not the case.  We want to summarize here a configuration that
has been presented in
\cite{BeCv,hep-th/0412250,hep-th/0308045} (for related discussions see also
\cite{Paris,BeJe,hep-th/0311146,hep-th/0411279,hep-th/0412250}).

In order to be consistent with the metric Ansatz and to
preserve the Poincar{\' e} symmetry in four dimensions, we allow for
(general) flux components in the internal space ${\cal Y}$ combined
with a 4-form flux proportional to external volume form yielding
a Freud-Rubin parameter $\lambda$:
\be062
\ba{l}
F^{(2)} = \frac{1 }{ 2} \, F_{mn}^{(2)} dy^m \wedge dy^n \quad , \qquad
H = \frac{1}{3} \, H_{mnp} dy^m \wedge dy^n \wedge dy^p \ , \\[2mm]
F^{(4)} = \lambda \, dx^0 \wedge dx^1 \wedge dx^2 \wedge dx^3
+ \frac{1}{4}\, F^{(4)}_{mnpq} \, dy^m \wedge dy^n \wedge dy^p \wedge dy^q \ .
\ea
\ee
Since the type IIA supergravity is non-chiral we can combine both
Majorana-Weyl spinors into one Majorana spinor and can take as
spinor Ansatz for an \N=1 vacuum (with SU(3) structures)
\be072
\epsilon =  (a \theta + b \theta^\star) \otimes \eta + cc 
\ = \ \theta \otimes ( a \eta + b^\star \eta^\star ) + cc \ .
\ee
There are two special cases: if $ab = 0$ the 4-dimensional spinor is
Weyl and if $b = a^\star$ we have a Majorana spinor.  The Killing
spinor equations (\ref{040}) can be solved by employing the relations
(\ref{230}) and one finds three solutions.

\noindent
{\em (i) If $\ \eta$ is Majorana-Weyl, ie.\ $ab=0$}

\noindent
For this spinor Ansatz, the mass and all RR-fields have to vanish:
\be773
F_{mn} = G_{mnpq} = W = dA = m=0 
\ee
and only the fields from the NS-sector are non-trivial
\be622
J \wedge H =  {^\star d \phi} \quad , \qquad H \wedge \Omega =0  \ .
\ee
One might have expected this result, because the NS-sector is common
to all string models, and a common solution can only be described by
one 10-dimensional Majorana-Weyl (Killing) spinor.  The non-zero
components of the 3-form $H$ can be decomposed with respect to SU(3)
representations, ie.\ we have a $(3,0),\ (2,1),\ (2,1)_0$ part, where
the subscript indicates the non-primitive part (ie.\ the components
encoded in $H \wedge J \neq 0$). The real forms are of course the real
and imaginary part of this projected forms. As solution one finds, the
holomorphic part $H^{(3,0)}$ has to vanish, $H^{(2,1)}_0$ fixes the
dilaton and if the primitive part $H^{(2,1)}$ is non-zero the internal
space is non-Kahlerian. A prototype example that solves these
equations is the NS5-brane supergravity solution, but there are also
other examples
\cite{NUPHA.B274.253,hep-th/0211118,BeBe,hep-th/0408121}.

\noindent
{\em (ii) If $\ \eta$ is Majorana, ie.\ $a = b^\star$ and $m=0$}

\noindent
This massless case can be lifted to 11-dimensional dimensions and the
only solution that one finds lifts the internal space to a
$G_2$-holonomy space, ie.\ only the 2-form is non-zero and the 4-form
is trivial.  This is the solution discussed in \cite{Paris} which
relates the warp factor and the 2-form by a monopole equation
\be928
\ba{l}
e^{A} d A =  - \frac{i }{8} \, e^\phi \, {^\star (F \wedge \bar \Omega)}
\quad ,\qquad \phi = - 3 U \ , \\[2mm]
0= F^{(4)} = H \wedge \Omega = F^{(2)} \wedge J \ .
\ea
\ee
Obviously, intersecting 6-branes is the prototype solution in this
class and the relation to $G_2$-holonomy spaces in M-theory without
any 4-form flux, identifies the moduli space of this configuration
with the moduli space of the corresponding 7-manifold and as before,
this flux background does not fix all moduli.

\noindent
{\em (iii) If $\ \eta$ is Majorana, ie.\ $a = b^\star$  but $m \neq 0$}

\noindent
This is the generic situation; also if $a \neq b^\star$, one gets an
equivalent solution. In this case all fluxes are non-zero and the
dilaton and warp factor are constant \cite{BeCv}
\bea671
&&0= d\phi = dA \quad  , \qquad 
H = H_0 \, {\rm Im}  \Omega \ ,
  \\ \label{672}
&& F^{(2)} =  F^{(2)}_0 \, J \quad    ,\qquad 
F^{(4)} = F^{(4)}_0\,  J \wedge J \ , \qquad
\eea  
with the coefficients given by
\be673
F_0^{(2)} = - \frac{2 }{ 9} \, \lambda \  , \quad
F_0^{(4)} = \frac{m }{ 20} \ , \quad
H_0 = -\frac{2m }{ 5} \, e^{\phi}  \ , \quad
\sqrt\frac{85}{ 2} \, \lambda  = 9 m \ .
\ee  
Hence, the dilaton is fixed by the ratio of the (quantized) fluxes
\be837
e^\phi = - \frac{ H_0 }{ 8\,  G_0} \ .
\ee
In this case the external space cannot be flat, but must be
anti-deSitter with a (negative) cosmological constant
\[
\Lambda = - e^{4\phi} \, m^2 = - \Big( \, \frac{H_0 }{ 8 G_0} \, \Big)^4 m^2
\ .
\]
The geometry of the internal space is nearly Kahler which is
equivalent to weak SU(3) holonomy and these spaces can be defined by
the differential equations
\[
dJ\  \sim \ {\rm Im}\; \Omega \quad , \qquad d\Omega \ \sim \ J \wedge J
\ .
\]
This means that for nearly Kahler spaces only the first torsion class
${\cal W}_1$ is non-trivial and moreover, the cone over any nearly
Kahler space gives a $G_2$ holonomy space \cite{710}. This can be used
to construct explicit examples.

The simplest spaces of weak SU(3) holonomy is ${\mathbb S}^6$ or
${\mathbb S}^3 \times {\mathbb S}^3$; other regular examples are the
twistor spaces over ${\mathbb S}^4$ or ${\mathbb{CP}}^2$.  These
spaces have no geometrical (closed string) moduli and therefore 
can serve as a starting point for exploring cosmological implications
as the KKLT scenario on the IIB side \cite{hep-th/0301240}.  In doing
this one has to wrap (anti) branes around supersymmetric (calibrated)
cycles and the most interesting candidates on the IIA side are wrapped
(anti) D6-branes, which can wrap cycles of the space ${\mathbb S}^3
\times {\mathbb S}^3$. The anti-D6-branes are of course a source for
(negative) RR-fields and to ensure (meta) stability the vacuum should
not contain the corresponding flux (ie.\ $dC_1 = 0$) so that these
anti-D6-branes cannot directly decay with the background flux.  Since
the vacuum has only a massive B-field flux, the anti-6-branes cannot
decay directly, but only due to the non-perturbative process where the
anti-D6-branes breaking-up leaving anti-NS5-branes at their
endpoints. These anti-5-branes can then decay with the NS-B-field
flux. Therefore, we should expect, that meta-stable deSitter vacua
should exist in the same way as on the IIB side. Moreover, since the
space ${\mathbb S}^3 \times {\mathbb S}^3$ has three supersymmetric
3-cycles, which intersect at an SU(3) angle, the intersection of
three anti-D6-branes supports chiral matter and if one wraps an equal
number of branes no orientifold projections are necessary
\cite{hep-th/0308046}.

We should add, that the solution presented here may have a
generalization, were further components of the 2-form are non-zero. As
shown in \cite{hep-th/0412250} additional non-primitive (1,1)
components of $F^{(2)}$ do not break supersymmetry and as consequence
the internal space is not nearly Kahler anymore. But unfortunately an
explicit example is not known yet.

%%%%%%%%%%%%%%%%%%%%%%%%%%%%%%%%%%%%%%%%%%%%%%%%%%%%%%%

\bigskip

\qquad {\bf Type IIB fluxes with SU(3) structure}

\medskip

%%%%%%%%%%%%%%%%%%%%%%%%%%%%%%%%%%%%%%%%%%%%%%%%%%%%%%%%%%

\noindent
The type IIB side has been discussed in the literature already
extensively, see
\cite{GrPo,hep-th/0107264,KST,hep-th/0212278,hep-th/0307142,
hep-th/0403220,hep-th/0404107} and we want to summarize here
only some aspects.  Again we admit only fluxes that are consistent
with 4-dimensional Poincar{\' e} symmetry, ie.\ they have components along
the internal space with the only exception of the 5-form, that has to
have components along the external space; required by the
self-duality.

An important property on the IIB side is, that as long as one keeps
SU(3) structures, the vacuum has to be flat, ie.\ a cosmological
constant can be generated by fluxes \cite{hep-th/0502154}. This may
indicate, that SU(3) structures always yield potentials of the
no-scale form which are positive definite, but this needs to be
verified for the most general fluxes consistent with SU(3)
structures. Recall, the no-scale structure is only an approximation
and corrections (quantum corrections, D3-instanton corrections etc.)
do not respect this property and yield anti deSitter vacua and
therefore these corrections have to break the SU(3) down to SU(2)
structures [cp.\ also the discussion after eq.\ (\ref{526})].  Let us
stress that we are using here only supersymmetry and therefore our
approach is valid for classical and quantum geometry as long as at
least four supercharges remain unbroken!

Depending on the concrete form of the spinor one finds again different
solutions and the most general spinor, consistent with SU(3)
structures can be written as
\be661
\epsilon = a \; [\theta \times \eta] + b^\star \; 
[\theta^\star \times \eta^\star]
\ee
where both spinors are chiral and $a$ and $b$ are complex coefficients;
we refer to \cite{hep-th/0307142,hep-th/0403220} for a classification
of the different spinor Ans\"atze.  In comparison to the IIA spinor
in (\ref{072}), this spinor is Weyl, but in general not Majorana --
only for $a=b$, the 10-dimensional spinor $\epsilon$ is Majorana-Weyl which
gives again the NS-sector solution that we encountered in $(i)$ on the
IIA side. There is another special sub-class of solution, namely if
$ab = 0$ which was explored to a large extend in \cite{GrPo,KST}.  The
special interest in this case comes due to the fact, that it still
allows the internal space to be Calabi-Yau and in the following we will
summarize this case as well as present the solution for the general
case, which has recently been found in \cite{hep-th/0502154}.

\bigskip

\noindent
{\em (i) Calabi-Yau flux compactifacations}

\noindent
Let us  summarize the solution in \cite{GrPo} and consider the spinor
\[
\epsilon = \theta \times \eta \ . 
\]
Using the chirality of $\eta$, both terms in the dilatino variation
$\delta \lambda$ in (\ref{006}) have to vanish separately, yielding
two equations
\be120 
G\, \eta = 0 \quad , \qquad P \, \eta^\star=0 
\ee
{from} which one infers, due to the relations (\ref{230}), that in
holomorphic coordinates: $ G_{abc} = G_{ab}{}^b = 0$ and $P$ is a
holomorphic vector. Similarly, in the gravitino variation, both terms
have to vanish separately yielding the constraint for the 3-form flux
\[
G \, \eta^\star = G_m \, \eta^\star =  0
\]
giving: $G_{\bar a \bar b \bar c}= G_{a}{}^{bc} \Omega_{bcd} = 0$ and
therefore the 3-form flux has to be primitive (ie.\ $G\wedge J =0$) and
of (2,1) type. The Ansatz for the (external) 5-form components are
\be002 
 F^{(ext)}_5 = 5\, e^{-4A} \,vol_4 \wedge dZ
\ee
and selfduality property fixes the remaining components.  Now, the gravitino
variation vanishes for a holomorphic $P$ only if
\be627
Z \sim e^{4A} \ .
\ee
Finally, the geometry is fixed by the differential equation obeyed by
Weyl spinor which becomes in this case
\be362
\nabla_m \eta = \frac{ i}{ 2} \, Q_m \, \eta
\ee
where $Q_m$ was defined in (\ref{032}). By inspecting the torsion
components in (\ref{526}), this spinor equation implies that only
${\cal W}_5\neq 0$ and all others vanish, which means that the space
is Kahler and $Q$ is the Kahler connection.  Since $Q$ is a specific
function on one complex coordinate $\tau$, only specific Kahler
geometries are possible, namely Kahler spaces related to wrapped
7-branes (which can be identified with singularities of the
holomorphic axion/dilaton).  If the vector $Q$ vanishes and the
axion-dilaton is trivial, the spinor is covariantly constant and hence
the space can have at most SU(3) holonomy and is Calabi-Yau.  Since
the corresponding potential has the no-scale structure, there is at
least one un-fixed modulus. This can only be fixed if one considers a
flux vacuum yielding a 4-dimensional anti-deSitter vacuum, which
however, was not compatible with SU(3) structures and one has instead
to consider SU(2) structures.  We have also to keep in mind, that
fluxes that generate a 4-dimensional cosmological constant will
always, if one includes the back reaction, render the internal space
to a non-Kahler geometry as it should be for any compactification that
fixes all moduli.

\bigskip

\noindent
{\em (ii) General flux compactifacations}

\noindent
Interestingly it is possible to solve the Killing spinor equations
without making any assumptions \cite{hep-th/0502154}. The type IIB
supergravity in our notation has a local U(1) symmetry which becomes
manifest if we define the phase $e^{2i\theta}=\frac{1+i\bar \tau }{
1-i\tau}$ and write the fields in (\ref{031}) as \cite{hep-th/9805131}
\be255 \ba{rcl} 
Q_M= \partial_M\theta -\frac{\partial_M\tau_1
  }{2\tau_2 } \quad ,\quad P_M=ie^{2i\theta}\frac{\partial_M\tau
  }{2\tau_2} \quad ,\quad
G_{(3)}=i\frac{e^{i\theta}}{\sqrt{\tau_2}}(d A_{(2)}-\tau dB_{(2)})\ .
\ea \ee
The phase $\theta$ drops out from the equations of motion as well as
from the Bianchi identities and the underlying symmetry is the local U(1)
gauge transformation
\be134
\epsilon \rightarrow e^{i g} \epsilon
\quad , \qquad 
\theta \rightarrow \theta + g 
\ee
for some function $g$.  This local symmetry is due to the coset
$SL(2,R)/U(1)$ which is parameterized by the scalar fields of type IIB
supergravity and implies that the phase $\theta$ can be chosen
freely, one can take $\theta =0$ (string theory convention) or
$e^{2i\theta}=\frac{1+i\bar
\tau }{ 1-i\tau}$ (supergravity convention) or any other value.
Recall, we are working in the Einstein frame which explains the
pre-factor $\tau_2^{-1/2} = e^{\phi/2}$ in the 3-form $G_3$.

We can write the spinor (\ref{661}) as
\be992
\epsilon=e^{\frac{A-i\omega}{ 2}} \, 
\Big(\sin\alpha\, [\zeta\otimes\chi] +\cos\alpha\, [
\zeta^\star\otimes\chi^\star] \, \Big)
\ee
where the appearance of the warp factor is a consequence of the
gravitino variation \cite{hep-th/0307142,hep-th/0502154}.  We absorbed
the common phase of $a$ and $b$ into the spinor ($\chi =e^{i\beta}
\chi_0$) and this phase drops out in most of the calculations.

The 5-form flux is again parameterized by the function $Z$ as in
(\ref{002}), and for the 3-form flux one find the form
\be047
G = \frac{1}{4} e^{-2A - i\omega} 
J \wedge \Big(\cot \alpha \; P_i dz^i + \tan \alpha 
\; P_{\bar i} d\bar z^i \Big) + G^{(prim)}
\ee
with the primitive part obeying: $J\wedge G^{(prim)} = \Omega \wedge
G^{(prim)} = \bar \Omega \wedge G^{(prim)} = 0$; $P_i$ is the
holomorphic part of the vector introduced in (\ref{255})
and $z^i$  denote the  three coordinates parameterizing the
internal space. Now, the solution of the Killing spinor equation is given 
in terms of one holomorphic function
\be991
f = f(z^i)
\ee
and can be written as \cite{hep-th/0502154}
\be997
\ba{rcl}
\tau &=& c_0 + i\, e^{-\phi_0}\, \frac{ |f|^2 \, \cos 2\alpha }{ 
f \, \sin^2\alpha   + f^\star \, \cos^2\alpha} \ , \\[2mm] 
e^{-4A}&=&  \frac{{\rm Re} f }{ 4 |f|^2}\, \frac{\sin^2 2\alpha }{\cos 2\alpha}
 \ ,
 \\[2mm]
Z &=& \frac{|f|^2 }{ {\rm Re} f} \, \frac{\cos^2 2\alpha }{ \sin^2 2\alpha} 
\\[2mm]
\tan(\theta+\omega) &=& - \frac{{\rm Im} f}{ {\rm Re} f} \cos 2\alpha \ .
\ea
\ee
Using the local symmetry (\ref{134}) we can set $\omega$ or $\theta$
to any fixed value, but the combination $\theta + \omega$ is gauge
invariant.  Note, supersymmetry leaves one function (in addition to
the holomorphic function $f$) free which has to be fixes by the
Bianchi identities or equations of motion; this is the master function
in \cite{hep-th/0403005}.  We chose here $\alpha$, which is the mixing
angle between the two spinors, but one may also take $Z$ which can be
fixed by the Bianchi identity $dF_5 \sim G \wedge \bar G$.  The
Calabi-Yau case is of course a special (where $\alpha \simeq 0$),
where the axion-dilaton and the vector $P$ are holomorphic.  For the
general case, the internal geometry is a complex manifold and becomes
(conformal) Kahler if: $(i)$ $G \wedge \bar G = 0$, which is ensured
if the primitive part of $G$ is of (2,1) type, and $(ii)$ $dZ \wedge
dA = 0$, which can be seen as a constraint on the function $f$.
Another special case are the solutions describing supergravity flows,
that correspond to the case where the holomorphic function is constant
$f = constant$. For more discussion we refer to \cite{hep-th/0502154}.

%%%%%%%%%%%%%%%%%%%%%%%%%%%%%%%%%%%%%%%%%%%%%%%%
%% BACKMATTER
%%%%%%%%%%%%%%%%%%%%%%%%%%%%%%%%%%%%%%%%%%%%%%%%

\begin{acknowledgement}
I would like to thank Gianguido Dall'Agata,
Hermann Nicolai and Kelly Stelle for numerous
fruitful discussions
\end{acknowledgement}

%% If the bibliography is produced without BibTeX comment out the

% \nocite{*}
% \bibliographystyle{aipproc}   % if natbib is available
% \bibliographystyle{utphys}   % if natbib is available
% \bibliographystyle{aipprocl} % if natbib is missing

%% \hyphenation{Post-Script Sprin-ger}
\providecommand{\href}[2]{#2}\begingroup\raggedright\endgroup

\end{document}